\documentclass[12pt,twoside]{article}
\usepackage{epsfig}
\usepackage[hmargin=25mm,vmargin=30mm,twoside]{geometry}

\usepackage{amssymb}
\usepackage{wasysym}
\usepackage{amsthm}
\usepackage{graphicx}
\usepackage{multido}
\def\R{{\mathbb R}}

\def\C{{\mathbb C}}

\def\kasten{$~~\mbox{\hfil\vrule height6pt width5pt depth-1pt}$ }
\def\dam{{\partial_\nu\partial^\nu}}
\def\pf{{\noindent \bf Proof. }}
\newtheorem{theorem}{Theorem}[section]

\newtheorem{definition}[theorem]{Definition}
\newtheorem{lemma}[theorem]{Lemma}

\begin{document}
\pagestyle{myheadings} \markboth{ H. Gottschalk}{ Scattering of dipole fields } \thispagestyle{empty}

\title{Scattering theory for dipole quantum fields}
 \author{Hanno Gottschalk}
\maketitle {\small

 \begin{center}
  Institut f\"ur angewandte Mathematik, Wegelerstr. 6, D-53115 Bonn, Germany\\
  gottscha@wiener.iam.uni-bonn.de
\end{center}
}

\begin{abstract}
In the present work a general frame for the scattering
theory of local, relativistic dipole quantum fields is presented and 
some models of interacting dipole fields are considered, i.e. local,
relativistic quantum fields with indefinite metric which
asymptotically do not converge to free fields, but to free dipole
fields. Also, we give explicit formulae for the (nontrivial)
scattering matrix of dipole in- and out- fields for these models. Furthermore we show how related
dipole degrees of freedom occur in the perturbation theory of certain two dimensional models, e.g. massive sine-Gordon or sinh-Gordon models.
\end{abstract}
{\small \noindent {\bf Keywords}: {\it Dipole fields, QFT with
indefinite metric, asymptotic states.}}

\section{Introduction}
In positive metric quantum field theory (QFT) the existence of
asymptotic states follows from the Wightman axioms and assumptions
on the mass spectrum of the two-point function (first and second
mass-gap), cf. \cite{Ha,RS,Ru}. The proof, however, relies
crucially on the positivity of Wightman functions.

The Wightman framework of local, relativistic quantum field theory
(QFT) turned out to be too narrow for theoretical physicists, who
were interested in handling situations involving in particular
gauge fields (like in quantum electrodynamics). For several
reasons which are intimately connected with the needs of the
standard procedure of the perturbative calculation of the
scattering matrix (for a detailed discussion, see \cite{Str}), the
concept of QFT with indefinite metric was introduced \cite{MS}.

The possible asymptotic behaviour of interacting quantum fields
with indefinite metric is not restricted to the usual free fields:
Other possible candidates are free dipole fields (and their
generalizations) which are local, relativistic quantum fields with
a genuinely indefinite metric on the space of states generated
from the vacuum. Such fields have been studied in a number of
articles, cf. \cite{US} and references therein.

The main motivation for the study of dipole quantum fields is threefold: (i) massless 
dipole fields in $d=4$ dimensions have logarithmic increase at spatially separated arguments that is
expected for fields that expose confinement \cite{US}; (ii) dipole degrees of freedom occur gauges beyond the Feynman gauge
in the Gupta-Bleuler Formalism \cite{Ste}; (iii) in the perturbation theory of certain models with exponential \cite{HK} or trigonometric \cite{AHK} interaction
dipole degrees of freedom occur in perturbation theory - although not present in the completely summed theory. To make sense of the
results of perturbation theory one requires a formalism that is capable to describe also the scattering of dipole fields.  This will be explained in section 6 of this work.   

In \cite{AGW4,BGL} local, relativistic and interacting quantum
fields have been constructed with indefinite metric state space.
These models generalize the class of models studied in
\cite{AGW1,AGW2,AGW3,AIK}. In \cite{AGW4} conditions have been
given, when such QFT models have a well-defined (and nontrivial)
scattering behaviour with the asymptotic fields given by usual
free fields. For these cases a general framework of scattering
theory has been proposed in \cite{AG}. On the other hand, there
are fields in the class studied in \cite{AGW4,BGL} which have
Haag-Ruelle like scattering amplitudes diverging polynomially in
time. The same occurs in the perturbation theory of exponential models in $d=2$.

In this article it is demonstrated that in the simplest
case this ill-defined scattering behaviour can be understood in
the sense that dipole degrees of freedom have been neglected:
Taking free dipole fields instead of usual free fields as
asymptotics of the interacting quantum  field models leads to a
well-defined scattering behaviour.

The article is organized as follows: In Section 2 dipole quantum fields are introduced. Then the scattering theory
withe dipole field as asymptotic fields is developed (Section 3). In Section 4 asymptotic states for dipole fields are being constructed via construction
of the form factor functional and an indfinite metric GNS construction and a criterium for the Morchio--Strocchi modified Wightman axioms.
In Section 5
we present local, relativistic models for quantum
fields with indefinite metric that do not have usual free fields as asymptotics, but dipole fields. In Section 6 we finally show how dipole
degrees of freedom occur in the first order perturbation theory of the exponential model. Some technicalities on the asymptotics of certain distributions are given in Appendix A.

\section{Free dipole fields}
Let $d\geq 2$ be the space-time dimension. A free dipole field of
mass $m> 0$ is a local, relativistic quantum field which fulfills
the equations of motion
\begin{equation}
(-\dam-m^2)^2\phi(x)=0
\end{equation}
(but $(-\dam-m^2)\phi(x)\not=0$ in contrast to the usual fee
field) and commutation relations
\begin{equation}
\left[ \phi(x),\phi(y)\right]_-={\alpha\over
i\pi}\Delta_m(x-y)+{\beta\over i\pi}\Delta'_m(x-y)
\end{equation}
$\beta \not=0$,
\begin{equation}
\Delta_m(x)={i\pi\over (2\pi)^{d/2}}\int_{\R^d}e^{ik\cdot
x}(\delta^+_m(k)-\delta^-_m(k))\, dk
\end{equation}
and
\begin{equation}
\Delta'_m(x)={i\pi\over (2\pi)^{d/2}}\int_{\R^d}e^{ik\cdot
x}(\delta^{'+}_m(k)-\delta^{'-}_m(k))\, dk
\end{equation}
with
\begin{equation}
\delta_m^\pm(k)=\theta(\pm
k^0)\delta(k^2-m^2)~~~\mbox{and}~~~~\delta_m^{'\pm}(k)=\theta(\pm
k^0)\delta'(k^2-m^2)\, .
\end{equation}

Since the two-point function does not admit a
K\"allen-Lehmann representation, dipole fields always are quantum
fields with indefinite metric \cite{MS,US}.

The equation of motion for dipole fields is a PDE of order 4,
hence the dipole field has an additional degree of freedom
compared with the usual free field ( note that the the spectral
condition rules out 2 degrees of freedom for the dipole field and
one degree of freedom for the usual free field. Degree of freedom
here means the numbers of functions which have to be specified on
a space-like hyperplane to determine the solution).

\section{The one particle wave operator}
In Haag-Ruelle theory \cite{Ha,Ru,He} asymptotic states are being
constructed by definition of the finite time wave operator on the
one particle space, which is then extended to the multi-particle
states by taking tensor products of the one particle wave
operator. By this strategy one circumvents the pitfalls of Haag's
theorem and Jost-Schroer theorem (cf. \cite{SW}) which say that
the interaction picture does not exist. With a similar strategy it
is also possible to construct asymptotic states in the case of QFT
with indefinite metric \cite{AG}.

But such asymptotics does not exist in all cases of local,
relativistic quantum fields with indefinite metric \cite{AGW4},
since scattering amplitudes of some of the models studied in that
reference diverge polynomially if one tries to calculate them on
the basis of the Haag-Ruelle one particle wave operator. As we
will demonstrate below, the asymptotics for such quantum fields is
rather given by a free dipole field (at least in the simplest
case) than by a usual free field. Thus, taking into account the
additional dipole degree of freedom, one can remove this
divergence and get a well-defined asymptotics. To carry out this
construction, we first have to find the right one particle wave
operator for the dipole field.

Such a one particle wave operator $\Omega^{d\,\rm in/out}_t$ is
required to fulfill the following conditions:
\begin{itemize}
\item[(i)] $\Omega^{d\,\rm in/out}_t:{\cal S}(\R^d)\to{\cal S}(\R^d)$
acts as a multiplication operator in energy-momentum space (in
analogy to the Haag-Ruelle wave operator).
\item[(ii)] $\Omega^{d\,\rm in/out}_t$ leaves free dipole fields invariant,
i.e. $\phi\circ\Omega^{d\,\rm in/out}(f)=\phi(f)\forall f\in{\cal
S}(\R^d), t\in\R$.
\item[(iii)] For the (interacting) current field $j(x)=(-\dam-m^2)\phi(x)$
 the action of $\Omega_t^{d\rm\, in/out}$ coincides with the action of the
usual Haag-Ruelle one-particle wave operator up to a term
proportional to $(-\dam-m^2)j(x)$ (i.e. up to a term which
vanishes in the free case).
\end{itemize}
Let $\varphi\in C^\infty_0(\R,\R)$ with support in
$(-\epsilon,\epsilon)$ where $0<\epsilon<m^2$ and
$\varphi(\kappa)=1$ if $\kappa \in(-\epsilon/2,\epsilon/2)$. We
define $\chi^\pm(k)=\theta(\pm k^0)\varphi(k^2-m^2)$ with $\theta$
the Heaviside step function and we set
\begin{equation}
\chi_{t}(a,k)=\left\{
 \begin{array}{ll}
\chi^+(k) e^{-i( k^0-\omega)t} +\chi^-(k)e^{-i( k^0+\omega)t} &
\mbox{$a=$in}\\ 1 & \mbox{$a=$loc}\\
 \chi^+(k) e^{i(
k^0-\omega)t} +\chi^-(k)e^{i( k^0+\omega)t} & \mbox{$a=$out}
\end{array}
\right.
\end{equation}
with $\omega=\sqrt{|{\bf k}|^2+m^2}$. We then define
\begin{equation}
\chi_t^d({\rm in/out},k)=\left[1\pm{it\over
2k^0}(k^2-m^2)\right]\chi_t({\rm in/out},k)
\end{equation}
with the $+$ sign belonging to "in" and the $-$ sign to "out" and
we set $\chi^d_t({\rm loc},k)=1$ . We finally define
\begin{equation}
\Omega_t^{d\,\rm in/out}={\cal F}\chi^d_t({\rm in/out},k)\bar{\cal
F} \, ,
\end{equation}
where ${\cal F}$ ($\bar {\cal F}$) stands for the (inverse)
Fourier transform.

 $\Omega^{d\rm in/out}_t$ fulfills the
above items (i) and (iii) by construction. Using Lemma \ref{A3lem}
and Lemma \ref{A4lem} of the appendix one can verify that also
(ii) holds.

 Our choice of $\Omega_t^{d\,\rm in/out}$ is not the only
possible one: in particular one can add terms proportional to
$p(t)e^{\pm i(k^0\pm\omega)t}(k^2-m^2)^2$ (with $p(t)$ a
polynomially bounded function) to $\chi^d_t({\rm in/out},k)$ and
one obtains another wave operator fulfilling items (i)-(iii) and
which gives the same scattering behaviour for the models studied
in Section 5. Such "higher order corrections" are needed if one
wants to generalize the possible asymptotic behaviour beyond free
dipole fields.

\section{Construction of asymptotic states}
Here we give a general scheme for the construction of asymptotic
states following \cite{AG}:

Let $\underline{\cal S}$ be the Borchers' algebra over ${\cal
S}(\R^d)$ and let $\underline{\cal S}^{\rm ext}$ be the
``extended'' Borchers' algebra over the test function space ${\cal
S}_1^{\rm ext}={\cal S}(\R^d,\C^3)$, which is the space of
Schwartz functions with values in $\C^3$. For $a=$in/loc/out we
define $J^a:{\cal S}_1\to{\cal S}_1^{\rm ext}$ to be the injection
of ${\cal S}_1$ into the first/second/third component of ${\cal
S}^{\rm ext}_1$, i.e. $J^{\rm in}f=(f,0,0),J^{\rm
loc}f=(0,f,0),J^{\rm out}f=(0,0,f)$, $f\in{\cal S}_1$. Then $J^a$
uniquely induces a continuous unital $*$-algebra homomorphism
$\underline{J}^a:\underline{\cal S}\to\underline{\cal S}^{\rm
ext}$ given by $\underline{J}^a=\oplus_{n=0}^\infty J^{a\otimes
n}$.

We also define a suitable ``projection''
$\underline{J}:\underline{\cal S}^{\rm ext}\to\underline{\cal S}$
as the unique continuous unital $*$-algebra homomorphism induced
by  $J:{\cal S}_1^{\rm ext}\to{\cal S}_1$, $J(f^{\rm in},f^{\rm
loc},f^{\rm out})=f^{\rm in}+f^{\rm loc}+f^{\rm out}$.

We then define $\Omega^d_t:{\cal S}^{\rm ext}_1\to{\cal S}_1^{\rm
ext}$ by
\begin{equation}
\label{3.10eqa} \Omega^d_t=\left(\begin{array}{ccc} \Omega_t^{d\,
\rm in} &0&0\\0&1&0\\0&0&\Omega^{d\,\rm out} _t\end{array}\right).
\end{equation}
Next, we introduce the multi parameter
$\underline{t}=(t_1,t_2,\ldots),
 t_n=(t_n^1,\ldots,t_n^n), t_n^l\in\R$ and we write $\underline{t}\to+ \infty$ if $t_n^l\to+\infty$
in any order, i.e. first one $t_n^l$ goes to infinity, then the
next etc. . We say that the limit $\underline{t}\to+\infty$ of any
given object exists, if it exists for $t_n^l\to+\infty$ in any
order and it does not depend on the order. We now  define the
finite times wave operator $\underline{\Omega}^d_{\underline
{t}}:\underline{\cal S}^{\rm ext}\to\underline{S}$ as
\begin{equation}
\label{3.11eqa}
\underline{\Omega}^d_{\underline{t}}=\underline{J}\circ\oplus_{n=0}^\infty\Omega^d_{n,t_n}~,~~~\Omega^d_{0,t_0}=1,~~
\Omega^d_{n,t_n}=\otimes_{l=1}^n\Omega^d_{t_n^l}.
\end{equation}
Furthermore, we define the finite times in- and out- wave
operators $\underline{\Omega}_{\underline{t}}^{d\,\rm
in/out}:\underline{\cal S}\to\underline{\cal S}$ as
$\underline{\Omega}^d_{\underline{t}}\circ \underline{J}^{\rm
in/out}$. Up to changes of the time parameter which do not matter
in the limit $\underline{t}\to+\infty$, the wave operators
$\underline{\Omega}^{d(\rm in/out)}_{\underline{t}}$ are
$*$-algebra homomorphisms, as can be easily verified from the
definitions.
\begin{definition}
\label{3.1def} {\rm (i) Let $\underline{W}\in\underline{\cal S}'$
be a Wightman functional s.t. the functionals
$\underline{W}\circ\underline{ \Omega}^d_{\underline{t}}$ converge
in $\underline{\cal S}^{\rm ext '}$ as $\underline{t}\to+\infty$.
We then define the dipole form factor functional
$\underline{F}^d\in
 \underline{\cal S}^{\rm ext '}$ associated to $\underline W$ as this limit, i.e.
\begin{equation}
\label{3.14eqa}
\underline{F}^d=\lim_{\underline{t}\to+\infty}\underline{W}\circ
\underline{\Omega}^d_{\underline{t}}.
\end{equation}
\noindent (ii) The scattering matrix $\underline{S}$ associated to
$\underline{W}$ is defined by
\begin{eqnarray}
\label{3.15eqa}
\underline{S}(\underline{f},\underline{g})&=&\underline{F}^d(\underline{J}^{\rm
in}\underline{f}\otimes \underline{J}^{\rm
out}\underline{g})\nonumber \\
&=&\lim_{\underline{t},\underline{t}'\to+\infty}\underline{W}(\underline{\Omega}_{\underline{t}}^{d\,\rm
in}\underline{f}\otimes\underline{\Omega}_{\underline{t}'}^{d\,\rm
out}\underline{g})~~\forall
\underline{f},\underline{g}\in\underline{\cal S}.
\end{eqnarray}
}
\end{definition}

We then get
\begin{theorem}
\label{3.1theo} We suppose that $\underline{W}\in\underline{\cal
S}'$ is the Wightman functional of a local, relativistic QFT with
indefininte metric \cite{Hoff2,MS}. We assume that the dipole form
factor functional $\underline{F}^d$ associated to $\underline{W}$
exists and fulfills a Hilbert space structure condition (cf.
\cite{Hoff2,MS}). Then there exists a  quantum field with
indefinite metric $\Phi$ (note that the in- loc- and out-
components of $\Phi$ in general are not local w.r.t. each other)
over ${\cal S}^{\rm ext}={\cal S}(\R^d,\C^3)$ such that the
relativistic quantum fields with indefinite metric $\phi^{\rm
in/loc/out}=\Phi\circ J^{\rm in/loc/out}$ over ${\cal S}(\R^d)$
s.t. $\phi=\phi^{\rm loc}$ fulfills the LSZ asymptotic condition
\cite{LSZ}, namely
\begin{equation}
\label{3.16eqa} \lim_{t\to+\infty}\phi({\Omega}^{d\,\rm
in/out}_{t}f)= \phi^{\rm in/out}(f) ~~\forall f\in{\cal S}(\R^d)
\end{equation}
\end{theorem}

The idea of the proof is to construct the weak asymptotic
limits of the vacuum expectation values first and to do a GNS-like
construction of the quantum field $\Phi$ with an indefinite metric
state space then. This automatically implies the asymptotic
condition (\ref{3.16eqa}). For the details see \cite{AG}.

So far there is no proof for the fact that the in- and out-
fields obtained in this way are free dipole fields. Such a general
proof requires the generalization of Theorem XI.111 of \cite{RS}
to the case of dipole fields - in particular we need fast
dispersion of functions $h^t(x)=\bar{\cal F}({1\over
2k^0}e^{i(k^0-\omega)t}(k^2-m^2)\chi^+(k)\hat f(k))$ in ${\bf x}$
as $t\to\infty$. $\|h^t(x^0,.)\|_{\bf x}\sim t^{-5/2}$ for a
suitable Schwartz norm $\|.\|_{\bf x}$ on ${\cal S}(\R^{d-1})$
is sufficient. This can be
obtained from the usual dispersion $\sim t^{-3/2}$ (for $d=4$) for
functions $h^t(x)=\bar{\cal F}({1\over
2k^0}e^{i(k^0-\omega)t}(k^0+\omega)\chi^+(k)\hat f(k))$ by
differentiation w.r.t. $t$. In the cases we study in the following
section one can prove that the asymptotic fields are free dipole
fields by explicit calculations.

\section{Models of interacting dipole fields}
Let $\psi(t)$, $t\in \R$, be an infinitely differentiable L\'evy
characteristic \cite{BF} and let $c=\linebreak-(d^2\psi(t)/ dt^2)|_{t=0}$.
Let $\eta(x)$ be a ultralocal "noise" field over the Euclidean
space-time $\R^d$ with functional Fourier transform (cf. e.g. \cite{It} for the definition of random fields via functional Fourier transforms)
\begin{equation}
\label{5.1eqa}
C_F(f)=\exp\left\{\int_{\R^d}\psi(f)-{c\over m^2}\langle\nabla f,\nabla
f\rangle-{c\over m^4}\langle\nabla f,(-\Delta+m^2)\nabla f\rangle
dx\right\}
\end{equation}
We consider the stochastic partial differential equation (SPDE)
\begin{equation}
\label{5.2eqa}
(-\Delta+m^2)^2\varphi=\eta
\end{equation}
which clearly is a Euclidean dipole type equation driven by the
noise field $\eta$. Using standard methods (cf. \cite{AGW1,AIK}) one
can solve this equation and calculate the moment functions of $X$:
\begin{equation}
\label{5.3eqa}
\left\langle\varphi(x_1)\cdots \varphi(x_n)\right\rangle=\sum_{I\in{\cal
P}^{(n)}}\prod_{\{j_1,\ldots,j_l\}\in I}
S_n^T(x_{j_1},\ldots,x_{j_l})
\end{equation}
where ${\cal P}^{(n)}$ is the collection of partitions of
$\{1,\ldots,n\}$ into disjoint subsets and the truncated Schwinger
functions $S_n^T$ are given by
\begin{equation}
\label{5.4eqa}
S_2^T(x,y)={c\over m^4}\,(-\Delta+m^2)^{-2}(x-y)
\end{equation}
and for $n\geq 3$
\begin{equation}
\label{5.5eqa}
S_n^T(x_1,\ldots,x_n)=c_n\int_{\R^d}\prod_{l=1}^n(-\Delta+m^2)^{-2}(x-x_l)\,dx
\end{equation}
where $c_n=i^{-n}(d^n\psi(t)/dt^n)|_{t=0}$. The analytic
continuation of these truncated Schwinger functions from imaginary
Euclidean time to real relativistic time has been performed in
\cite{AGW4}. We get for the Fourier transforms of the truncated
Wightman functions
\begin{equation}
\label{5.6eqa}
\hat W_2^T(k_1,k_2)=-{c\over m^4}\,\delta
^{'-}_m(k_1)\delta(k_1+k_2)
\end{equation}
and
\begin{equation}
\label{5.7eqa}
\hat W_n^T(k_1,\ldots,k_n)=(-1)^n\tilde c_n\left\{\sum_{j=1}^n
\prod_{l=1}^{j-1}\delta^{'-}_m(k_l)\,{1\over
(k^2-m^2)^2}\,\prod_{l=j+1}^n\delta_m^{'+}(k_l)\right\}\delta(\sum_{l=1}^nk_l)
\end{equation}
with $\tilde c_n=(2\pi)^{d(n-2)-2\over 2}c_n$.

 $1/(k^2-m^2)^2$ has to be understood in the sense of  Cauchy's principal value.

 It has been proven in \cite{AGW4} that the
 truncated Wightman functions $W_n^T=\bar{\cal
F}(\hat W_n^T)$ are the truncated vacuum expectation values of a
local, relativistic QFT with indefinite metric.

In the same reference it has been proven that the
usual form factor functional of the associated Wightman functional
does not exist since Haag-Ruelle like scattering amplitudes
diverge poynomially in time. Here we show that the asymptotic
behaviour of the given QFT with indefinite metric is rather given
by free dipole fields.

Using Lemma \ref{A4lem} and \ref{A6lem} one gets for $n\geq 3$
that
\begin{equation}
\lim_{t_n^1,\ldots,t_n^n\to+\infty}\prod_{l=1}^n\chi^d_{t_n^l}(a_l,k_l)\hat
W_n^T(k_1,\ldots,k_n)=\hat
F_n^{d,T(a_1,\ldots,a_n)}(k_1,\ldots,k_n),
\end{equation}
where
\begin{equation}
\hat F_2^{d,T(a_1,a_2)}(k_1,k_2)=\hat W_2^T(k_1,k_2)
\end{equation}
and for $n\geq 3$
\begin{eqnarray}
 &&\hat F_{n}^{d,T(a_1,\ldots,a_n)}(k_1,\ldots,k_n)\nonumber\\
&=&\left\{\sum_{j=1}^n\prod_{l=1}^{j-1} \delta^{'-}_{m}(k_l) \hat
\Delta'_{m}(a_j,k_j)\prod_{l=j+1}^n\delta^{'+}_{m}(k_l)\right\}\delta
(\sum_{l=1}^nk_l),
\end{eqnarray}
($a_l=$in/loc/out), and
\begin{equation}  \hat \Delta_{m}' (a,k)=\left\{
\begin{array}{ll}
 i\pi (\delta_{m}^{'+}(k)-\delta_{m}^{'-}(k)) & \mbox{ for $a=$in}\\
1 \left/(k^2-m^2)^2\right.                & \mbox{ for $a=$loc}\\
-i\pi (\delta_{m}^{'+}(k)-\delta_{m}^{'-}(k)) & \mbox{ for
$a=$out}
\end{array}\right.
\end{equation}

We summarize our discussion with the following theorem:

\begin{theorem}
Let $\underline{W}$ be the Wightman functional associated to the
Fourier transformed, truncated Wightman functions defined above.
Then

\noindent (i) $\underline{W}$ fulfills the modified Wightman
axioms of Morchio and Strocchi \cite{MS}.

\noindent (ii) The usual form factor functional for
$\underline{W}$ does not exist, but the dipole form factor
functional exists and fulfills a Hilbert space structure
condition.

\noindent (iii) Thus, there exist a local, relativistic quantum
field $\phi(x)$ with indefinite metric such that the local quantum
field converges to asymptotic fields $\lim_{t\to\infty}\phi\circ
\Omega_t^{d\,\rm in/out}(f)=\phi^{\rm in/out}(f)\forall f\in{\cal
S}(\R^d)$.

\noindent (iv) The asymptotic fields $\phi^{\rm in/out}(x)$
fulfill the equation of motion and commutation relations of the
free dipole fields.

\noindent (v) The S-matrix associated to the quantum field
$\phi(x)$ is given by
\begin{eqnarray}
&&\left\langle \phi^{\rm in}(k_r)\cdots\phi^{\rm in}(k_1)\Psi_0,
\phi^{\rm out}(k_{r+1})\cdots\phi^{\rm
out}(k_n)\Psi_0\right\rangle ^T\nonumber\\ &=&2\pi i \tilde c_n
\prod_{l=1}^n \delta^{'+}_m(k_l) \,
\delta(\sum_{l=1}^rk_l-\sum_{l=r+1}^nk_l),
\end{eqnarray}
where $k_1^0,\ldots,k_n^0>0$ (i.e. all $\phi^{\rm in/out}(k_l)$
are creation operators), $\Psi_0$ is the vacuum state and $n\geq
3$.
\end{theorem}
\pf (i) and (ii): This has been established in the above
discussion, that the form factor functional fulfills the Hilbert
space structure condition follows from a slight generalization of
Lemma 4.9 of \cite{AG}.

(iii) is a consequence of Theorem \ref{3.1theo}.

(iv) and (v) can be verified by calculations using the explicit
formulae for the truncated form factor functional (for similar
calculations in a related case however with usual free asymptotic
fields cf. \cite{AG}) . \kasten

 We note that there are strong indications that
for fields with a usual (non dipole) form factor functional
$\underline{F}$ \cite{AG} also the dipole form factor functional
$\underline{F}^d$ exists and that $\underline{F}=\underline{F}^d$
in this case. Using the Lemmas \ref{A3lem} and \ref{A5lem} below,
this e.g. can be verified for the class of models studied in
\cite{AG}.

\section{Dipoles in perturbation theory}

In this section we biefly show, how dipole degrees of freedom occur in the low order perturbation theory for the exponential \cite{HK} or trigonometric \cite{AHK} model in $d=2$. 
Although these models after analytic continuation fulfill the Wightman axioms and thus have a well-defined scattering theory without any dipoles showing up,
already in the first order contribution to the Wightman function we do get a dipole term which, however, does not contribute to the non-trivial part of the scattering amplitude if dipole asymtotics is taken into account.
If not, the scattering behavior of the first order contribution is ill defined.  

Let $\langle A\rangle=\int_{{\cal S}'(\R^2)} A(\varphi)d\mu_0(\varphi)$ be the expected value of $A$ with respect to the free field measure $\mu_0$ on ${\cal S}'(\R^2)$, i.e. the measure with characteristic functional 
\begin{equation}
{\cal C}(f)=\exp\left\{-\frac{1}{2}(f,(-\Delta+m^2)^{-1}f)\right\},~f\in{\cal S}(\R^2).
\end{equation}

 It is well-known that for $\rho$ a probability measure with support in $(-\sqrt{4\pi},\sqrt{4\pi})$ and $v(t)=\int_\R e^{\alpha t} \,d\rho(\alpha)$, 
the exponential interaction $\int_\Lambda :v(\varphi):\, dx$ for $\Lambda\subseteq \R^2$ compact is a positive function and square integrable with respect to $\mu_0$. We define the Schwinger functions of the interacting meausre with infra red cut-off $\Lambda$ as
\begin{equation}
\label{6.1eqa}
\left\langle\varphi(x_1)\cdots\varphi(x_n)\right\rangle_{\Lambda,\lambda}=\frac{\left\langle\varphi(x_1)\cdots\varphi(x_n)e^{-\lambda\int_\Lambda :v(\varphi):\, dx}\right\rangle}{\left\langle e^{-\lambda\int_\Lambda :v(\varphi):\, dx}\right\rangle}~,~~\lambda\geq0.
\end{equation} 
Since the potential term is square intrgrable, the Schwinger functions for $\lambda\geq0$ are twice differentiable (at $\lambda=0$ differentiable from the right) which gives us the first order expansion together with a control of the error due to Taylors lemma. Using also $(a-\lambda b)/(1-\lambda c)=a-\lambda b+\lambda ac+o(\lambda^2)$ and $\langle:\exp\alpha\varphi(x):\rangle=1$, we obtain 
\begin{eqnarray}
\label{6.2eqa}
\left\langle\varphi(x_1)\cdots\varphi(x_n)\right\rangle_{\Lambda,\lambda}&=&\left\langle\varphi(x_1)\cdots\varphi(x_n)\right\rangle\nonumber\\
&-&\frac{\lambda}{2}\int_\R\int_\Lambda\left\langle\varphi(x_1)\cdots\varphi(x_n):\exp\alpha\varphi(x):\right\rangle \, dxd\rho(\alpha)\nonumber\\
&+& \frac{\lambda}{2}\langle\varphi(x_1)\cdots\varphi(x_n)\rangle |\Lambda|+o(\lambda^2),
\end{eqnarray}
where $|\Lambda|$ is the volume of $\Lambda$. By a standard calculation
\begin{equation}
\label{6.3eqa}
\left\langle :\exp\alpha\varphi(x):e^{i\varphi(f)}\right\rangle=\exp\left[i\alpha(-\Delta+m^2)^{-1}f(x)-\frac{1}{2} (f,(-\Delta+m^2)^{-1}f)\right].
\end{equation}
Performing functional derivatives with respect to $f$ at $f=0$ and multiplying by $(-i)^n$, one obtains
\begin{eqnarray}
\label{6.4eqa}
\left\langle\varphi(x_1)\cdots\varphi(x_n):\exp\alpha\varphi(x):\right\rangle&=&\sum_{S\subseteq \{1,\ldots,n\}}\alpha^{|S|} \prod_{j\in S}(-\Delta+m^2)^{-1}(x_j-x)\\
&&~~~~~~~\times	\sum_{I\in {\cal P}_2(\{1,\ldots,n\}\setminus S)}\prod_{\{j,l\}\in I}(-\Delta+m^2)^{-1}(x_j-x_l)\nonumber
\end{eqnarray}
Here ${\cal P}_2(A)$ is the collection of all pair partitions of the set $A$. $|S|$ gives the cardinaliy of $S$.

Let us first consider the contribution to (\ref{6.2eqa}) that stems from those terms in 
(\ref{6.4eqa}) with $S=\emptyset$. Appearently those terms give a contriution $-\frac{\lambda}{2}|\Lambda|\langle\varphi(x_1)\cdots\varphi(x_n)\rangle$ that cancels the term in the last line of that equation. Removing the infra red cut-off via $\Lambda\nearrow \R^2$, one obtains
\begin{eqnarray}
\label{6.5eqa}
\left\langle\varphi(x_1)\cdots\varphi(x_n)\right\rangle_{\lambda}&=&\sum_{\emptyset\not=S\subseteq \{1,\ldots,n\}}\int_\R\alpha^{|S|}d\rho(\alpha) \int_{\R^2}\prod_{j\in S}(-\Delta+m^2)^{-1}(x_j-x)\, dx\\
&&~~~~~~~~~~\times	\sum_{I\in {\cal P}_2(\{1,\ldots,n\}\setminus S)}\prod_{\{j,l\}\in I}(-\Delta+m^2)^{-1}(x_j-x_l)+o(\lambda^2)\nonumber
\end{eqnarray}
For $S=\{s_1,\ldots,s_q\}$ and $I$ fixed, we can now do the analytic continuations factor by factor. The Fourier transformed analytic continuation of the pairing terms just gives the usual to point function $\delta_m^-(k_j)\delta(k_j+k_l)$ $(j<l)$. The first factor is a constant for $q=1$. For $q=2$, $S=\{s_1,s_2\}$ we get (\ref{5.4eqa}) up to a constant and hence the Fourier transformed analytic continuation up to a constant factor is given by (\ref{5.6eqa}), which is a dipole two point function. If $q\geq 3$, by \cite{AGW1} one obtains as the Fourier transformed analytic continuation
\begin{equation}
\label{6.6eqa}
\tilde c_{q}\left\{\sum_{j=1}^{q}
\prod_{l=1}^{j-1}\delta^{-}_m(k_{s_l})\,{1\over
(k_{s_j}^2-m^2)}\,\prod_{l=j+1}^{q}\delta_m^{+}(k_{s_l})\right\}\delta(\sum_{l=1}^{q}k_{s_l}).
\end{equation}  
Here $c_n$ in the definition of $\tilde c_n$ is replaced by the $n$-th moment of $\rho$. Appearently, the dipole degrees of freedom here do not contribute to scattering, if the proper dipole asymptotics is taken into account. Following the argument of section 5, one can show that the contribution of (\ref{6.6eqa}) to the dipole scattering amplitude is (see also \cite{AG})
\begin{equation}
\label{6.7eqa}
2\pi i \tilde c_{q}
\prod_{l=1}^q \delta^{+}_m(k_{s_l}) \,
\delta(\sum_{l=1}^rk_{s_l}-\sum_{l=r+1}^qk_{s_l}),
\end{equation}
i.e. the non trivial scattering concerns the classicle particle degrees of freedom, only. 

In fact, (\ref{6.7eqa}) is just what one would expect from the expansion $:\exp \alpha\varphi(x):= \sum_{l=0}^\infty\frac{\alpha^l}{l!}\linebreak :\varphi^l(x):$ with the first order term cancelled by a symmetric choice of $\rho$ ($\rho=\frac{1}{2}(\delta_{-1}+\delta_1)$ for massive $\sinh$-Gordon) and the second order term absorbed into a re-definition of the the mass, which however in the rigorous formulation of the exponential model is not needed. In fact, in that expansion any $:\varphi^l:$ interaction term to first order corresponds to a pure s-wave scattering (in terms of Feynman graphs: a star graph with $l$ legs) which is just (\ref{6.7eqa}). We have thus seen that
the dipole formulation of scattering allows us to get the correct 1st order scattering amplitude without any unnecessary re-definition of the mass. 

Here the calculations have been carried out for the exponential model. For the trigonometric model the argument is analogous with the only difference that $\alpha$ is replaced by $i\alpha$.
\appendix

\section{Some distributional identities}
Let $m>0$, $d\geq 2$, $\partial_0=(\partial/\partial k^0)$,
$\omega=\sqrt{|{\bf k}|^2+m^2}$.
\begin{lemma}
\label{A1lem} $\delta^{'\pm}_m(k)={1\over
2k^0}\partial_0\delta^{\pm}_m(k)\, .$
\end{lemma}
\pf By change of variables.\kasten
\begin{lemma}
\label{A2lem} $(k^2-m^2)\delta^{'\pm}_m(k)=-\delta^\pm_m(k)$.
\end{lemma}
\pf By Lemma \ref{A1lem}
 \begin{eqnarray*}
\int_{\R^d}(k^2-m^2)\delta^{'\pm}_m(k)\hat f(k)\,
dk&=&-\int_{\R^d}\delta^{\pm}_m(k)\,\partial_0\left[{(k^2-m^2)\over
2k^0}\hat f(k)\right] dk\\
&=&-\int_{\R^d}\delta^{\pm}_m(k)\left[{\partial_0(k^2-m^2)\over
2k^0}\right] \hat f(k)\, dk
\end{eqnarray*}
Since $\left[{\partial_0(k^2-m^2)\over
2k^0}\right]_{k^0=\pm\omega}=1$, the claim follows.
  \kasten
\begin{lemma}
\label{A3lem} $\chi^d_t(a,k)\delta^\pm_m(k)=\delta_m^\pm(k)$.
\end{lemma}
\pf Let $a=$ out ($a=$loc is trivial and $a=$in can be treated
analogously), then
 \begin{eqnarray*}
 \int_{\R^d}\delta^{\pm}_m(k)\chi^d_t({\rm out},k)\hat f(k)\, dk
 &=&\int_{\R^d}\delta_m^\pm(k)e^{i(k^0\mp\omega)t}\\
 &\times&\left[ 1-{it\over
 2k^0}(k^2-m^2)\right]\chi^\pm(k)\hat f(k)\, dk\\
 &=&\int_{\R^d}\delta^\pm_m(k)\hat f(k)\, dk
 \end{eqnarray*}
 \kasten
\begin{lemma}
\label{A4lem}
$\chi^d_t(a,k)\delta^{'\pm}_m(k)=\delta^{'\pm}_m(k)$.
\end{lemma}
\pf As in Lemma \ref{A3lem}, it suffices to consider $a=$out:
\begin{eqnarray*}
&&\int_{\R^d} \delta_m^{'\pm}(k) \chi^d_t({\rm out},k)\hat f(k)\,
dk\\ &=&\int_{\R^d}\delta^{'\pm}_m(k)e^{i(k^0\mp\omega)t}\left[
1-{it\over 2k^0} (k^2-m^2)\right]\chi^\pm(k)\hat f(k)\, dk\\ &=&
-\int_{\R^d}\delta^\pm_m(k) {it\over 2k^0} \left[1-{it\over
2k^0}(k^2-m^2)\right] \chi^{\pm}(k)\hat f(k)\, dk\\ &+&\int_{\R^d}
\delta_m^{'\pm}(k)\left[1-{it\over
2k^0}(k^2-m^2)\right]\chi^\pm(k)\hat f(k)\, dk\\
&=&\int_{\R^d}\delta^{'\pm}_m(k)\hat f(k)\, dk
\end{eqnarray*}
Here we used Lemma \ref{A2lem} in the last step. \kasten
\begin{lemma}
\label{A5lem} $\lim_{t\to+\infty}{\chi^d_t(a,k)\over
(k^2-m^2)}=\hat\Delta_m(a,k)$ (for the definition of $\hat
\Delta_m(a,k)$ cf. \cite{AG}).
\end{lemma}
\pf Let $a=$in/out ($a=$loc is trivial). Then
 \begin{eqnarray*}
 \lim_{t\to+\infty}\int_{\R^d}{{\chi}^d_t(a,k)\over(k^2-m^2)}\hat
 f(k)\, dk &=& \lim_{t\to\infty}\Bigg[ \int_{\R^d}
 {\chi_t(a,k)\over(k^2-m^2)}\hat f(k)\, dk\\
 &-&it\int_{\R^d} {\chi_t(a,k)\over 2k^0}\hat f(k)\, dk\Bigg]
 \end{eqnarray*}
 The second integral on the right hand side vanishes faster than
 any inverse polynomial in $t$ as $t\to+\infty$, thus the
 statement follows from Lemma 4.8 of \cite{AG}.
\kasten
\begin{lemma}
\label{A6lem} $\lim_{t\to+\infty}{\chi^d_t(a,k)\over
(k^2-m^2)^2}=\hat\Delta'_m(a,k)$.
\end{lemma}
\pf For $a=$out we calculate ($a=$ in can be treated analogously
and $a=$loc is trivial)
 \begin{eqnarray*}
&&\int_{\R^d}{\chi^d_t({\rm out},k)\over (k^2-m^2)^2}\hat f(k)\,
dk\\ &=& \int_{\R^d} {1\over k^2-m^2}\,
\partial_0\left[{1-{it\over 2k^0}(k^2-m^2)\over
2k^0}\chi_t({\rm out},k)\hat f(k)\right] dk\\ &=&
\int_{\R^d}{\chi_t({\rm out},k)\over (k^2-m^2)}\, \partial_0\left[
{1-{it\over 2k^0}(k^2-m^2)\over 2k^0} \hat f(k) \right]dk\\ &+&
it\int_{\R^d}{\chi_t({\rm out},k)\over (k^2-m^2)}\, {1-{it\over
2k^0}(k^2-m^2)\over 2k^0}\,\hat f(k)\, dk\\ &+& \int_{\R^d} {e^{
i(k^0-\omega)t}\partial_0\chi^+(k)+e^{i(k^0+\omega)t}\partial_0\chi^-(k)\over
(k^2-m^2)}\\ &\times& {1-{it\over 2k^0}(k^2-m^2)\over 2k^0}\, \hat
f(k)\, dk
 \end{eqnarray*}
 The third integral on the right hand side vanishes faster than any inverse polynomial as $t\to\infty$. For the first
 two integrals we get
 \begin{eqnarray*}
 &&\int_{\R^d} {\chi_t({\rm out},k)\over (k^2-m^2)}\,
 \partial_0\left[{1\over 2k^0}\,\hat f(k)\right]dk\\
 &+& it\int_{\R^d} {\chi_t({\rm out},k)\over (k^2-m^2)}{1\over 2k^0}\left[
 1-{\partial_0 (k^2-m^2)\over 2k^0}\right] \hat f(k)\, dk\\
 &-& it \int_{\R^d}\chi_t({\rm out},k)\,\partial_0 \left[{1\over
 (2k^0)^2}\,\hat f(k)\right] dk\\
 &+&t^2\int_{\R^d}\chi_t({\rm out},k)\,{1\over (2k^0)^2}\,\hat f(k)\, dk\, .
 \end{eqnarray*}
The last two integrals on the right hand side vanish faster than
any inverse polynomial as $t\to\infty$. This holds also true for
the second integral, since by Lemma 4.8 of \cite{AG} the limit of
the integral is
 $$ \int_{\R^d}{\hat \Delta_m({\rm out},k)\over
2k^0}\left[1-{\partial_0(k^2-m^2)\over 2k^0}\right] \hat f(k) \,
dk=0 $$ and furthermore
 $$ {d\over dt}\int_{\R^d}
{\chi_t({\rm out},k)\over (k^2-m^2)}{1\over 2k^0}\left[
 1-{\partial_0 (k^2-m^2)\over 2k^0}\right] \hat f(k)\, dk\to 0
 $$
faster than any inverse polynomial as $t\to+\infty$. Now the claim
follows from Lemma 4.8 of \cite{AG} and integration by parts.
 \kasten

\

\noindent {\bf Acknowledgements.} I gratefully acknowledge
valuable discussions with S. Albeverio, F. Strocchi, J.-L. Wu and G. Lechner.

\end{document}